\documentclass[a4paper, 12pt, oneside]{article}
\usepackage[cp1251]{inputenc}
\usepackage[english, russian]{babel}
\usepackage{graphics,amsfonts,amsmath,bm,amssymb,array,eucal}
\usepackage[dvips]{graphicx}
\usepackage[flushleft,small]{caption2}
\renewcommand{\baselinestretch}{1.15}

\begin{document}
\thispagestyle{empty} \large
\renewcommand{\abstractname}{\, }

 \begin{center}
\bf The Kramers Problem for Quantum Fermi Gases with Velocity - Dependent
Collision Frequency and Specular - Diffusive Boundary Conditions
\end{center}\medskip
\begin{center}
  \bf A. Yu. Kvashnin\footnote{$malin_85@mail.ru$},
  A. V. Latyshev\footnote{$avlatyshev@mail.ru$} and
  A. A. Yushkanov\footnote{$yushkanov@inbox.ru$}
\end{center}\medskip

\begin{center}
{\it Faculty of Physics and Mathematics,\\ Moscow State Regional
University, 105005,\\ Moscow, Radio str., 10--A}
\end{center}\medskip

\begin{abstract}
    The classical Kramers problem of the kinetic theory is solved.
The Kramers problem about isothermal sliding for quantum Fermi
gases is considered. Quantum gases with the velocity - dependent
collision frequency are considered. Specular - diffusive boundary 
conditions are applied. Dependence of isothermal sliding on the 
resulted chemical potential is found out.


\end{abstract}

\tableofcontents
\setcounter{secnumdepth}{4}

\addcontentsline{toc}{section}{Введение}
\markboth{}{ВВЕДЕНИЕ}
\begin{center}
\item{}\section*{Введение}
\end{center}

В первой половине прошлого столения Юлинг и Уленбек \cite{141} 
обобщили кинетическое уравнение Больцмана на случай
квантовых газов. 

Поведение квантовых газов представляет значительный интерес и в настоящее 
время  \cite{121} и \cite{129}.

Первые аналитические решения граничных задач для модельных
кинетических уравнений для квантовых газов с постоянной частотой
столкновений были получены в работах 
\cite{63}--\cite{69}. 

Аналитические решения граничных задач для квантовых газов
с переменной частотой столкновений получены в работах 
\cite{118a}--\cite{KLY-1}.

Вопросы скольжения одноатомных классических газов вдоль плоской 
твердой поверхности изложены в монографиях \cite{91}--\cite{93}.

Укажем на другие работы по скольжению классических газов вдоль плоской 
твердой поверхности \cite{33}--\cite{79}.

Начало аналитическим методам в кинетической теории было положено работой
К. Черчиньяни \cite{105}. В его работах (и вместе с соавторами) 
\cite{106}--\cite{110} были заложены идеи формулировки и решения граничных
задач и с новыми (более адекватными) граничными условиями и с частотой
столкновений, зависящей от модуля скорости молекул.

Следует отметить еще ряд работ, посвященных аналитическому решению 
граничных задач кинетической теории \cite{113}--\cite{144}, а также численному
и приближенному решению таких задач \cite{132}--\cite{136}.

Настоящая работа является продолжением нашей предыдущей работы \cite{KLY-1},
где была рассмотрена классическая задача кинетической теории -- задача
Крамерса об изотермическом скольжении для квантовых ферми--газов 
с переменной частотой столкновений и с 
полностью диффузными граничными условиями. 

В настоящей работе эта задача решается с более общими зеркально -- 
диффузными граничными условиями. При этом применяется метод, изложенный
в ряде наших работ (см., например, \cite{83}--\cite{83b}). Показано, что
применяемый метод обладает высокой эффективностью. Так, уже второе приближение
дает ошибку, менее 0.01\%.

\begin{center}
\item{}\section{Формулировка граничных условий
и постановка задачи}
\end{center}

\markboth{}{ГРАНИЧНЫЕ УСЛОВИЯ И ПОСТАНОВКА ЗАДАЧИ}

В качестве кинетического уравнения будем использовать линейное уравнение
(5.11) из \cite{KLY-1}
$$
\mu\dfrac{\partial h}{\partial x_1}+h(x_1,\mu)=\dfrac{3}{4}
\int\limits_{-1}^{1}(1-{\mu'}^2)h(x_1,\mu')d\mu'.
\eqno{(1.1)}
$$

Уравнение (1.1) выведено из нелинейного кинетического уравнения для 
квантовых ферми--газов, которое в условиях задачи Крамерса может быть
записано в виде:
$$
v_x\dfrac{\partial f}
{\partial x}= \nu(\mathbf{v})[f_F^*(x,\mathbf{v},t)-
f(x,\mathbf{v},t)].
\eqno{(1.2)}
$$

При этом функция распределения $f(x,\mathbf{v},t)$ связана с функцией
$h(x_1,\mu)$ равенством:
$$
f(x,\mathbf{v},t)=f_F(v)+g(v)C_yh(x_1,\mu),
$$
где
$$
g(v)=\dfrac{\exp(\beta v^2-{\mu}/{kT})}{\big(1+
\exp(\beta v^2-{\mu}/{kT})\big)^2}, \quad \beta=\dfrac{m}{2kT},\quad
\mathbf{C}=\sqrt{\beta}\mathbf{v},
$$
$$
f_F(v)=\dfrac{1}{1+\exp\big(\beta v^2-{\mu}/{kT}\big)},\quad
\mu=\dfrac{C_x}{C},
$$
$x_1=x/(\nu_0 \sqrt{\beta})$ -- безразмерная координата, $\nu_0=1/\tau$,
$\tau$ -- время между двумя последовательными столкновениями ферми--частиц газа,
$l=\tau v_T$ -- средняя длина свободного пробега ферми--частиц,
$v_T=1/\sqrt{\beta}$ -- тепловая скорость ферми--частиц, 
$\nu(\mathbf{v})=v/l$.

Сформулируем зеркально -- диффузные граничные условия для
функции распределения:
$$
f(+0, \mathbf{v})=qf_F(v)+(1-q)f(+0,-v_x,v_y, v_z), \quad
v_x>0,
$$
где $q$ -- коэффициент диффузности, $0 \leqslant q \leqslant 1$,
$f_F(v)$ -- абсолютный фермиан.

Параметр $q$ -- часть молекул, рассеивающихся границей диффузно,
т.е. уходящие
от стенки молекулы имеют максвелловское распределение по
скоростям, $1-q$ -- часть молекул, рассеивающихся зеркально.

Продолжим функцию распределения на сопряженное полупространство
симметричным образом:
$$
f(x,\mathbf{v})=f(-x, -v_x,v_y,v_z).
\eqno{(1.3)}
$$

Продолжение на полупространство $x<0$ позволяет включить
граничные условия в уравнения задачи.

Такое продолжение функции распределения
позволяет фактически рассматривать две задачи, одна из которых
определена в "положительном"\, полупространстве $x>0$, вторая --
в отрицательном "полупространстве"\, $x<0$.

Сформулируем зеркально -- диффузные граничные условия для
функции распределения соответственно для "положительного"\, и
для "отрицательного"\, полупространств:
$$
f(+0, \mathbf{v})=qf_F(v)+(1-q)f(+0,-v_x,v_y, v_z), \quad
v_x>0,
\eqno{(1.4)}
$$
$$
f(-0, \mathbf{v})=qf_F(v)+(1-q)f(-0, -v_x,v_y, v_z).
v_x<0.
\eqno{(1.5)}
$$

Далее безразмерную координату $x_1$ снова будем обозначать через
$x$. Тогда для функции $h(x,\mu)$ вместо (1.3) мы имеем:
$$
h(x,\mu)=h(-x,-\mu), \qquad \mu>0.
$$

Граничные условия (1.4) и (1.5) преобразуются следующим образом:
$$
h(+0,\mu)=(1-q)h(+0,-\mu), \quad 0<\mu<1,
\eqno{(1.6)}
$$
$$
h(x,\mu)=h_{as}(x,\mu)+o(1), \quad x\to \infty,
\eqno{(1.7)}
$$
где
$$
h_{as}=2U_{sl}+2G_v(x-\mu),
$$
$U_{sl}= \sqrt{\beta}u_{sl}$ - искомая скорость изотермического
скольжения (безразмерная).

Правая часть уравнения (1.1) есть удвоенная массовая скорость
классического одноатомного газа \cite{LY-2}:
$$
U(x)=\dfrac{3}{8}\int\limits_{-1}^{1}(1-{\mu'}^2)h(x,\mu')d\mu'.
\eqno{(1.8)}
$$

Требуется найти скорость скольжения газа $u_{sl}$ и построить
функцию $h(x,\mu)$.
    Для решения этой задачи развивается новый
эффективный метод решения граничных задач с зеркально --
диффузными граничными условиями.

В основе предлагаемого метода лежит идея включить граничное
условие на функцию распределения в виде источника в кинетическое
уравнение.

Суть предлагаемого метода состоит в следующем. Сначала
формулируется в полупространcтве $x>0$ классическая задача
Крамерса об изотермическом скольжении с зеркально -- диффузными
граничными условиями. Затем функция распределения продолжается в
сопряженное полупространство $x<0$ четным образом по
пространственной и по скоростной переменным. В полупространстве
$x<0$ также формулируется задача Крамерса.

После того как получено линеаризованное кинетическое уравнение
разобьем искомую функцию (которую также будем называть функцией
распределения) на два слагаемых: чепмен --- энскоговскую функцию
распределения $h_{as}(x,\mu)$ и вторую часть функции
распределения $h_c(x,\mu)$, отвечающей непрерывному спектру:
$$
h(x,\mu)=h_{as}(x,\mu)+h_c(x,\mu),
$$ ($as \equiv asymptotic,
c\equiv continuous$).

В силу того, что чепмен -- энскоговская функция распределения
есть линейная комбинация дискретных решений исходного уравнения,
функция $h_c(x,\mu)$ также является решением кинетического
уравнения. Функция $h_c(x,\mu)$ обращается в нуль вдали от
стенки. На стенке эта функция удовлетворяет зеркально --
диффузному граничному условию.

Далее преобразуем кинетическое уравнение для функции
$h_c(x,\mu)$, включив в это уравнение в виде члена типа
источника, лежащего в плоскости $x=0$, граничное условие на
стенке для функции $h_c(x,\mu)$. Подчеркнем, что функция
$h_c(x,\mu)$ удовлетворяет полученному кинетическому уравнению в
обеих сопряженных полупространствах $x<0$ и $x>0$.

Это кинетическое уравнение мы решаем во втором и четвертом
квадрантах фазовой плоскости $(x,\mu)$ как линейное
дифференциальное уравнение первого порядка, считая известным
функцию $U_c(x)$. Из полученных решений находим
граничные значения неизвестной функции $h^{\pm}(x,\mu)$ при
$x=\pm 0$, входящие в кинетическое уравнение.

Теперь мы разлагаем в интегралы Фурье неизвестную функцию
$h_c(x,\mu)$, неизвестную функцию $U_c(x)$ и дельта --
функцию Дирака. Граничные значения неизвестной функции
$h_c^{\pm}(0,\mu)$ при этом выражаются одним и тем же интегралом
на спектральную плотность $E(k)$ функции $U_c(x)$.

Подстановка интегралов Фурье в кинетическое уравнение и
выражение функции $U_c(x)$ приводит к характеристической
системе уравнений. Если исключить из этой системы спектральную
плотность $\Phi(k,\mu)$ функции $h_c(x,\mu)$, мы получим
интегральное уравнение Фредгольма второго рода.

Считая градиент функции $U_c(x)$ заданным, разложим неизвестную
скорость скольжения, а также спектральные плотности функции $U_c(x)$
и функции распределения в ряды по степеням коэффициента
диффузности $q$ (это ряды Неймана). На этом пути мы получаем
счетную систему зацепленных уравнений на коэффициенты рядов для
спектральных плотностей. При этом все уравнения на коэффициенты
спектральной плотности для функции $U_c(x)$ имеют особенность
(полюс второго порядка в нуле). Исключая эти особенности
последовательно, мы построим все члены ряда для скорости
скольжения, а также ряды для спектральных плотностей функции $U_c(x)$
и функции распределения.

\begin{center}
  \item{}\section{Неоднородное кинетическое уравнение}
\end{center}

\markboth{ГЛАВА 3}{НЕОДНОРОДНОЕ КИНЕТИЧЕСКОЕ УРАВНЕНИЕ}

В этом п. мы преобразуем кинетическое уравнение для функции
$h_c(x,\mu)$, включив в это уравнение в виде члена типа
источника, лежащего в плоскости $x=0$, граничное условие на
стенке для функции $h_c(x,\mu)$. Функция
$h_c(x,\mu)$ удовлетворяет полученному кинетическому уравнению в
обеих сопряженных полупространствах $x<0$ и $x>0$.
Заметим, что функция $h_{as}(x,\mu)$ является решением уравнения (1.1).
Поэтому новая неизвестная функция
$h_c(x,\mu)=h(x,\mu)-h_{as}(x,\mu)$ удовлетворяет также
уравнению (1.1).

Так как вдали от стенки ($x\to \infty$) функция распределения
$h(x,\mu)$ переходит в чепмен --- энскоговскую $h_{as}(x,\mu)$,
то для функции $h_c(x,\mu)$, отвечающей непрерывному спектру,
получаем следующее граничное условие: $ h_c(\pm \infty,\mu)=0.
$
Отсюда для функции $U_c(x)$ получаем:
$$
U_c(\pm \infty)=0. 
$$

Граничные условия переходят в следующие:
$$
h_c(+0,\mu)=-h_{as}^+(+0,\mu)+$$$$+(1-q)h_{as}^+(+0,-\mu)+
(1-q)h_c(+0,-\mu), \quad \mu>0,
$$
$$
h_c(-0,\mu)=
-h_{as}^-(-0,\mu)+$$$$+(1-q)h_{as}^-(-0,-\mu)(1-q)h_c(-0,-\mu), \quad \mu<0.
$$

Перепишем эти условия в виде:
$$
h_c(+0,\mu)=h_0^+(\mu)+(1-q)h_c(+0,-\mu), \quad \mu>0,
$$
$$
h_c(-0,\mu)=h_0^-(\mu)+(1-q)h_c(-0,-\mu), \quad \mu<0,
$$
где
$$
h_0^{\pm}(\mu)=-h_{as}^{\pm}(0,\mu)+(1-q)h_{as}^{\pm}(0,-\mu)=
$$$$=
-2qU_{sl}(q)+(2-q)2G_v|\mu|.
$$

Учитывая симметричное продолжение функции распределения, имеем
$$ h_c(-0,-\mu)=h_c(+0,+\mu),\qquad
h_c(+0,-\mu)=h_c(-0,+\mu).
$$
Следовательно, граничные условия перепишутся в виде:
$$
h_c(+0,\mu)=h_0^+(\mu)+(1-q)h_c(-0,\mu), \quad \mu>0,
\eqno{(2.1)}
$$
$$
h_c(-0,\mu)=h_0^-(\mu)+(1-q)h_c(+0,\mu), \quad \mu<0.
\eqno{(2.2)}
$$

Граничные условия обеих задач можно объединить следующим
образом:
$$
h_c(\pm 0, \mu)=h_0^{\pm}(\mu)+(1-q)h_c(\mp 0,\mu), \quad \pm
\mu>0\quad (|\mu|<1),
\eqno{(2.3)}
$$
$$
h_c(\pm \infty,\mu)=0, \qquad \pm \mu<0 \quad(|\mu|<1).
\eqno{(2.4)}
$$

Здесь
$$
h_0^{\pm}(\mu)=-2qU_{sl}+2(2-q)G_v^{\pm }\mu,
$$
причем
$$
G_v^-=-G_v^+.
$$

Возьмем уравнение, содержащее оба граничные условия (2.3):
$$
\mu\dfrac{\partial h_c}{\partial x}+h_c(x,\mu)=$$$$=2U_c(x)+
|\mu|[h_0^{\pm}(\mu)-qh_c(\mp 0,\mu)]\delta(x),
\eqno{(2.5)}
$$
где $\delta(x)$ -- дельта--функция Дирака, $U_c(x)$ -- часть
массовой скорости классического газа, отвечающая непрерывному спектру,
$$
2U_c(x)=\dfrac{3}{4}\int\limits_{-1}^{1}(1-{\mu'}^2)h_c(x,\mu')d\mu'.
\eqno{(2.6)}
$$

Проверим, что граничные условия включены в уравнение (2.5).
Пусть, например, $\mu\in (0,1)$. Проинтегрируем обе части
уравнения (2.5) по $x$ от $-\varepsilon$ до $+\varepsilon$.
Получаем, что
$$
h_c(+\varepsilon,\mu)-h_c(-\varepsilon,\mu)=h_0^+(\mu)-
qh_c(-0,\mu).
$$

Переходя к пределу при $\varepsilon\to 0$ в этом выражении,
получаем в точности условие (2.3).

На основании определения массовой скорости (2.6) заключаем, что
для неё выполняется условие $U_c (\pm \infty)$=0. Следовательно,
в полупространстве $x>0$ профиль массовой скорости газа
вычисляется по формуле:
$$
U(x)=U_{as}(x)+\dfrac{3}{8}\int\limits_{-1}^{1}(1-{\mu'}^2)
h_c(x,\mu')d\mu',
$$
где
$$
U_{as}(x)=U_{sl}(q)+g_vx,
$$
а вдали от стенки имеет следующее линейное распределение:
$$
U(x)=U_{as}(x), \qquad x\to +\infty. 
$$

\begin{center}
\item{}\section{ Характеристическая система и интегральное
уравнение Фредгольма}
\end{center}

\markboth{ГЛАВА 3}{ХАРАКТЕРИСТИЧЕСКАЯ СИСТЕМА И УРАВНЕНИЕ ФРЕДГОЛЬМА}

Решая уравнение (2.5) при $x>0,\mu<0$, считая известной
массовую скорость $U_c(x)$, получаем, удовлетворяя граничным
условиям (2.3) и (2.4), следующее решение:
$$
h_c^+(x,\mu)=-\dfrac{\exp(-x/\mu)}{\mu}\int\limits_{x}^{+\infty}
\exp(t/\mu)2U_c(t)dt.
\eqno{(3.1)}
$$

Аналогично при $x<0,\mu>0$ находим:
$$
h_c^-(x,\mu)=\dfrac{\exp(-x/\mu)}{\mu}\int\limits_{-\infty}^{x}
\exp(t/\mu)2U_c(t)dt.
\eqno{(3.2)}
$$

Теперь уравнения (2.5) можно переписать, заменив второй член в
квадратной скобке из (2.5) согласно (3.1) и (3.2), в виде:
$$
\mu\dfrac{\partial h_c}{\partial x}+h_c(x,\mu)=
$$
$$=
2U_c(x)+ |\mu|[h_0^{\pm}(\mu)-qh_c^{\pm}(0,\mu)]\delta(x).
\eqno{(3.3)}
$$

В равенстве (3.3) граничные значения $h_c^{\pm}(0,\mu)$
выражаются через составляющую массовой скорости, отвечающей
непрерывному спектру:
$$
h_{c}^{\pm}(0,\mu)=-\dfrac{1}{\mu}e^{-x/\mu}
\int\limits_{0}^{\pm \infty}e^{t/\mu}2U_c(t)dt.
$$

Решение уравнения (2.5) и (2.6) ищем в виде интегралов Фурье:
$$
2U_c(x)=\dfrac{1}{2\pi}\int\limits_{-\infty}^{\infty}\exp(ikx)E(k)dk,
\quad
\delta(x)=\dfrac{1}{2\pi}\int\limits_{-\infty}^{\infty}\exp(ikx)dk,
\eqno{(3.4)}
$$
и
$$
h_c(x,\mu)=\dfrac{1}{2\pi}\int\limits_{-\infty}^{\infty}\exp(ikx)
\Phi(k,\mu)dk.
\eqno{(3.5)}
$$

При этом функция $h_c^+(x,\mu)$ выражается через спектральную
плотность $E(k)$ массовой скорости следующим образом:
$$
h_c^+(x,\mu)=-\dfrac{\exp(-x/\mu)}{2\pi\mu}\int\limits_{x}^{\infty}
\exp(t/\mu)dt\int \limits_{-\infty}^{\infty}\exp(ikt)E(k)dk=
$$
$$
=\dfrac{1}{2\pi}\int\limits_{-\infty}^{\infty}\dfrac{\exp(ikx)E(k)dk}
{1+ik\mu}.
$$

Аналогично,
$$
h_c^-(x,\mu)=\dfrac{1}{2\pi}\int\limits_{-\infty}^{\infty}
\dfrac{\exp(ikx)E(k)dk}{1+ik\mu}.
$$

Таким образом, последние два равенства можно объединить в одно:
$$
h_c(x,\mu)=\dfrac{1}{2\pi}\int\limits_{-\infty}^{\infty}
\dfrac{\exp(ikx)E(k)dk}{1+ik\mu}.
$$

Используя четность $E(k)$ можно записать:
$$
h_c^{\pm}(0,\mu)=\dfrac{1}{2\pi}\int\limits_{-\infty}^{\infty}
\dfrac{E(k)dk}{1+k^2\mu^2}=\dfrac{1}{\pi}\int\limits_{0}^{\infty}
\dfrac{E(k)dk}{1+k^2\mu^2}.
$$

Подставляя последнее равенство вместе с интегралами Фурье (3.4)
и (3.5) в уравнения (2.9) и (2.6), приходим к
характеристической системе:
$$
E(k)=\dfrac{3}{4}\int\limits_{-1}^{1}(1-\mu^2)\Phi(k,\mu)d\mu,
\eqno{(3.6)}
$$
$$
(1+ik\mu)\Phi(k,\mu)=
$$
$$
=E(k)-2qU_{sl}|\mu|+
2(2-q)G_v\mu^2-\dfrac{q|\mu|}{\pi}\int\limits_{0}^{\infty}
\dfrac{E(k_1)dk_1}{1+k_1^2\mu^2}.
\eqno{(3.7)}
$$

Подставляя (3.7) в (3.6), приходим к характеристическому
интегральному уравнению:
$$
E(k)L(k)+\dfrac{q}{\pi}\int\limits_{0}^{\infty}J_1(k,k_1)E(k_1)dk_1=$$$$=
-2qU_{sl}T_1(k)+2(2-q)G_vT_2(k),
\eqno{(3.8)}
$$
где
$$
J_n(k.k_1)=\dfrac{3}{2}\int\limits_{0}^{1}\dfrac{t^n(1-t^2)dt}
{(1+k^2t^2)(1+k_1^2t^2)},\quad n=0,1,2,\cdots,
\eqno{(3.9)}
$$
Уравнение (3.8) есть интегральное уравнение Фредгольма второго
рода.
$$
T_n(k)=J_n(k,0)=
\dfrac{3}{2}\int\limits_{0}^{1}\dfrac{t^n(1-t^2)dt}{1+k^2t^2},
$$
Нетрудно заметить,что
$$
L(k)=1-T_0(k)=1-\dfrac{3}{2}\int\limits_{0}^{1}\dfrac{1-t^2}{1+k^2t^2}dt=
k^2T_2(k).
$$

Решение уравнения (3.9) будем искать в виде
$$
E(k)=2(2-q)G_v\Big[E_0(k)+q\,E_1(k)+q^2\,E_2(k)+\cdots\big],
\eqno{(3.10)}
$$
$$
\Phi(k,\mu)=2(2-q)G_v\Big[
\Phi_0(k,\mu)+q\Phi_1(k,\mu)+q^2\Phi_2(k,\mu)+\cdots\Big].
\eqno{(3.11)}
$$

Скорость скольжения $U_{sl}(q)$ при этом будем искать в виде
$$
U_{sl}=\dfrac{2-q}{q}G_{v}
\Big[V_0+V_1q+V_2q^2+\cdots+V_nq^n+\cdots\Big].
\eqno{(3.12)}
$$
Подставим ряды (3.10)--(3.12) в уравнения (3.7) и (3.8).
Теперь эти интегральные уравнения распадаются на
эквивалентную бесконечную систему уравнений. В нулевом
приближении получаем следующую систему уравнений:
$$
E_0(k)L(k)=-V_0T_1(k)+T_2(k),
\eqno{(3.13)}
$$
$$
\Phi_0(k,\mu)(1+ik\mu)=E_0(k)+\Big[|\mu|-V_0\Big]|\mu|,
\eqno{(3.14)}
$$
В первом приближении:
$$
E_1(k)L(k)=-V_1T_1(k)- \dfrac{1}{\pi}\int\limits_{0}^{\infty}
J(k,k_1)E_0(k_1)dk_1,
\eqno{(3.15)}
$$
$$
\Phi_1(k,\mu)(1+ik\mu)=E_1(k)-V_1|\mu|-\dfrac{|\mu|}{\pi}
\int\limits_{0}^{\infty}\dfrac{E_0(k_1)dk_1}{1+k_1^2\mu^2}.
\eqno{(3.16)}
$$
Во втором приближении имеем:
$$
E_2(k)L(k)=-V_2T_1(k)- \dfrac{1}{\pi}\int\limits_{0}^{\infty}
J(k,k_1)E_1(k_1)dk_1,
\eqno{(3.17)}
$$
$$
\Phi_2(k,\mu)(1+ik\mu)=E_2(k)-V_2|\mu|-\dfrac{|\mu|}{\pi}
\int\limits_{0}^{\infty}\dfrac{E_1(k_1)dk_1}{1+k_1^2\mu^2}.
\eqno{(3.18)}
$$
В n-ом приближении получаем:

$$
E_n(k)L(k)=-V_nT_1(k)-
\dfrac{1}{\pi}\int\limits_{0}^{\infty}J_1(k,k_1)E_{n-1}(k_1)dk_1,
\eqno{(3.19)}
$$

$$
\Phi_n(k,\mu)(1+ik\mu)=$$$$=E_n(k)-V_n|\mu|- \dfrac{|\mu|}{\pi}
\int\limits_{0}^{\infty}\dfrac{E_{n-1}(k_1)dk_1}{1+k_1^2\mu^2}.
 \; n=1,2,\cdots.
\eqno{(3.20)}
$$

\begin{center}
  \item{}\section{Решение задачи}
\end{center}

\begin{center}
  \item{}\subsection{Нулевое приближение}
\end{center}

Из формулы (3.13) для нулевого приближения находим:
$$
E_0(k)=\dfrac{T_2(k)-V_0T_1(k)}{L(k)}.
\eqno{(4.1)}
$$
Нулевое приближение массовой скорости на основании (4.1) равно:
$$
U_c^{(0)}(x)=G_v\dfrac{2-q}{2\pi}\int\limits_{-\infty}^{\infty}
\exp(ikx)E_0(k)\,dk=$$$$=G_v\dfrac{2-q}{2\pi}
\int\limits_{-\infty}^{\infty}
\exp(ikx)\dfrac{-V_0T_1(k)+T_2(k)}{L(k)}dk.
\eqno{(4.2)}
$$
Согласно (4.2) наложим на нулевое приближение массовой скорости
требование: $U_c(+\infty)=0$. Это условие приводит к тому, что
подынтегральное выражение из интеграла Фурье (4.2) в точке
$k=0$ конечно. Следовательно, мы должны устранить полюс второго
порядка в точке $k=0$ у функции $E_0(k)$. Замечая, что
$$
T_2(0)=\dfrac{3}{2} \int\limits_{0}^{1} t^2(1-t^2)dt=\dfrac{3}
{15}, \qquad T_1(0)=\dfrac{3}{2} \int\limits_{0}^{1}
t(1-t^2)dt=\dfrac{3}{8},
$$
находим нулевое приближение $V_0$:
$$
V_0=\dfrac{T_2(0)}{T_1(0)}=\dfrac{8}{15}.
\eqno{(4.3)}
$$

С помощью (4.3) найдём, опуская вычисления, числитель выражения
(4.1):
$$
T_2(k)-\dfrac{8}{15}T_1(k)=
k^2\Big(\dfrac{8}{15}T_3(k)-T_4(k)\Big).
$$
Таким образом,
$$
T_2(k)-\dfrac{8}{15}T_1(k)=k^2\varphi_0(k),
$$
где
$$
\varphi_0(k)=\dfrac{3}{2}\int\limits_{0}^{1}\dfrac{(1-t^2)t^3(\frac{8}{15}-t)}
{1+t^2k^2}dt=\Big(\dfrac{8}{15}T_3(k)-T_4(k)\Big).
$$
Итак, возвращаясь к (4.1), имеем:
$$
E_0(k)=\dfrac{\varphi_0(k)}{T_2(k)}.
$$

Согласно (3.5) и (3.14) находим:
$$
h_c^{(0)}(x,\mu)=\dfrac{1}{2\pi}\int\limits_{-\infty}^{\infty}
\Big[E_0(k)+|\mu|(|\mu|-V_0)\Big] \dfrac{\exp(ikx)dk}{1+ik\mu}.
$$

\begin{center}
  \item{}\subsection{Первое приближение}
\end{center}

Перейдем к первому приближению. В первом приближении из
уравнения (3.15) находим:
$$
E_1(k)=-\dfrac{1}{L(k)}\Big[V_1T_1(k)+\dfrac{1}{\pi}
\int\limits_{0}^{\infty}
\dfrac{J(k,k_1)}{T_2(k_1)}\varphi_0(k_1)dk_1\Big].
\eqno{(4.4)}
$$

Первая поправка к массовой скорости имеет вид
$$
U_c^{(1)}(x)=G_v\dfrac{2-q}{2\pi}\int\limits_{-\infty}^{\infty}
\exp(ikx)E_1(k)\,dk.
$$

Требование $U_c(+\infty)=0$ приводит к требованию конечности
подынтегрального выражения в предыдущем интеграле Фурье.
Устраняя полюс второго порядка в точке $k=0$, находим:
$$
V_1=-\dfrac{1}{T_1(0)\pi}\int\limits_{0}^{\infty}
T_1(k_1)E_0(k_1)dk_1=0.0518.
\eqno{(4.5)}
$$

Используя соотношение (3.15) и непосредственно проверяемое равенство
$$
T_n(k)=T_n(0)-k^2T_{n+2}(k),
$$
находим, что
$$
E_1(k)=\dfrac{1}{T_2(k)\pi}\int\limits_{0}^{\infty}S(k,k_1)E_0(k_1)dk_1,
\eqno{(4.6)}
$$
где
$$
S(k,k_1)=k_1^2\Big[\dfrac{T_3(k)T_3(k_1)}{T_1(0)}-J_5(k,k_1)\Big],
$$
или, кратко:
$$
E_1(k_1)=\dfrac{\varphi_1(k_1)}{T_2(k_1)},\quad
\text{где}\quad \varphi_1(k_1)=\dfrac{1}{\pi}\int\limits_{0}^{1}
\dfrac{S(k_1,k_2)}{T_2(k_2)}\varphi_0(k_2)\,dk_2,
$$
или
$$
\varphi_1(k_1)=\dfrac{1}{\pi}\int\limits_{0}^{1}
S(k_1,k_2)E_0(k_2)\,dk_2.
$$
Теперь подставляя (4.6) в (3.16) находим первое приближение
спектральной плотности функции распределения:
$$
\Phi_1(k,\mu)=\dfrac{1}{1+ik\mu}\Big[E_1(k)
-V_1|\mu|-\dfrac{|\mu|}{\pi}
\int\limits_{0}^{\infty}\dfrac{E_0(k_1)\,dk_1}
{1+k_1^2\mu^2}\Big].
$$

\begin{center}
  \item{}\subsection{Второе приближение}
\end{center}

Перейдем ко второму приближению задачи -- это уравнения (3.17) и
(3.18). Из уравнения (3.17) находим:
$$
E_2(k)=-\dfrac{1}{L(k)}\Big[V_2T_1(k)+\dfrac{1}{\pi}
\int\limits_{0}^{\infty} J(k,k_1)E_1(k_1)\,dk_1\Big].
\eqno{(4.7)}
$$

Вторая поправка к массовой скорости имеет вид:
$$
U_c^{(2)}(x)=G_v\dfrac{2-q}{2\pi}\int\limits_{-\infty}^{\infty}
e^{ikx}E_2(k)\,dk.
$$

Условие $U_c(+\infty)=0$ приводит к требованию ограниченности
функции $E_2(k)$ в точке $k=0$. Устраняя полюс второго порядка в
точке $k=0$ в правой части равенства для $E_2(k)$, находим:

$$
V_2=-\dfrac{1}{T_1(0)\pi}\int\limits_{0}^{\infty}T_1(k_1)E_1(k_1)dk_1=
-0.0031,
\eqno{(4.8)}
$$
и, кроме того,
$$
E_2(k)=\dfrac{1}{T_2(k)\pi}\int\limits_{0}^{\infty}S(k,k_1)E_1(k_1)dk_1.
\eqno{(4.9)}
$$
Для второго приближения спектральной плотности функции
распределения из уравнения (3.18) получаем:
$$
\Phi_2(k,\mu)=\dfrac{1}{1+ik\mu}\Bigg[E_2(k)-V_2|\mu|
-\dfrac{|\mu|}{\pi} \int\limits_{0}^{\infty}\dfrac{E_1(k_1)dk_1}
{1+k_1^2\mu^2}\Bigg].
$$

\begin{center}
  \item{}\subsection{Высшие приближения}
\end{center}

В третьем приближении из уравнения (3.19) получаем:
$$
E_3(k)=-\dfrac{1}{L(k)}\Big[V_3T_1(k)+\dfrac{1}{\pi}
\int\limits_{0}^{\infty}J(k,k_1)E_2(k_1)dk_1\Big].
$$
Как и ранее, устраняя полюс второго порядка в точке $k=0$,
получаем:
$$
V_3=-\dfrac{1}{\pi}\int\limits_{0}^{\infty}J(0,k_1)E_2(k_1)dk_1=
-\dfrac{1}{\pi}\int\limits_{0}^{\infty}T_1(k_1)E_2(k_1)dk_1,
$$

Проводя аналогичные рассуждения, для $n$--го приближения
согласно (3.19) и (3.20) получаем (при $n=1,2, \cdots$):
$$
V_n=-\dfrac{1}{\pi
T_1(0)}\int\limits_{0}^{\infty}T_1(k)E_{n-1}(k)\,dk,\quad
\text{где}\quad E_n(k)=\dfrac{\varphi_n(k)}{T_2(k)},
$$
$$
\Phi_n(k,\mu)=\dfrac{1}{1+ik\mu}\Bigg[E_n(k)-V_n|\mu|-
\dfrac{|\mu|}{\pi}
\int\limits_{0}^{\infty}\dfrac{E_{n-1}(k_1)dk_1}
{1+k_1^2\mu^2}\Bigg].
$$

\begin{center}
  \item{}\section{Сравнение с точным решением и профиль массовой
скорости}
\end{center}

\markboth{ГЛАВА 3}{СРАВНЕНИЕ С ТОЧНЫМ РЕШЕНИЕМ И МАССОВАЯ СКОРОСТЬ}

\begin{center}
  \item{}\subsection{Сравнение с точным решением}
\end{center}

Таким образом, безразмерная скорость изотермического скольжения
нами найдена в виде ряда:
$$
U_{sl}=G_v \dfrac{2-q}{q}[V_0+V_1q+V_2q^2+...].
$$

Сравним нулевое, первое и второе приближения при $q=1$ с точным
решением.
$$
U_{sl}=G_v \dfrac{2-q}{q}[V_0+V_1q+V_2q^2+...V_nq^n], \qquad
n=0,1,2...
$$

Точное значение скорости скольжения в случае диффузного
рассеяния таково:
$$
U_{sl}(1)=0.5819G_v.
$$

Нетрудно проверить, что в нулевом (максвелловском) приближении
$$
U_0^{sl}(1)=0.533(3)G_v,
$$
т.е. нулевое приближение дает ошибку $8.4\%$.

В первом приближении получаем
$$
U_1^{sl}(1)=(V_0+V_1)G_v=0.5851G_v,
$$
это значит, что первое приближение дает ошибку $0.5\%$.

Во втором приближении
$$
U_2^{sl}(1)=(V_0+V_1+V_2)G_v=0.5820G_v,
$$
т.е. второе приближение дает ошибку $0.01\%$.

Переходя, как и ранее, к размерной скорости скольжения, получаем:
$$
u_{sl}=\dfrac{V_1}{\sqrt{\beta}}G_v,\quad\text{где}\quad
g_v=\nu_0 G_v.
$$
Отсюда имеем:
$$
u_{sl}=\dfrac{V_1}{\sqrt{\beta}l}G_vl= \dfrac{V_1}{\sqrt{\beta}l
\nu_0}g_vl.
$$

Выбирая длину свободного пробега согласно Черчиньяни \cite{94}, как
$l=\dfrac{\eta}{\rho}\sqrt{\pi\beta} $, получаем
следующее соотношение:
$$
l \sqrt{\beta}=\dfrac{\eta}{\rho}\sqrt{\pi \beta}\sqrt{\beta}=
\dfrac{8\sqrt{\pi}l_1(\alpha)}{15\nu_0 l_0(\alpha)}.
$$
Возвращаясь к выражению для размерной скорости скольжения,
учитывая полученное, приходим к следующему выражению:
$$
u_{sl}=\dfrac{15V_1\nu_0l_0(\alpha)}{8 \sqrt{\pi}
l_1(\alpha)\nu_0}lg_v.
$$
В итоге получаем следующую формулу для размерной скорости
скольжения газа
$$
u_{sl}=K_v(\alpha,q)l g_v,
\eqno{(5.1)}
$$
где $K_v$ -- коэффициент изотермического скольжения, $l$ --
длина свободного пробега,
$$
K_v(\alpha,q)=
\dfrac{15l_0(\alpha)}{8\sqrt{\pi}l_1(\alpha)}
\dfrac{2-q}{2}\Big[V_0+V_1q+V_2q^2+...\Big].
$$

\begin{center}
  \item{}\subsection{Профиль скорости газа в полупространстве и ее значение
  у стенки}
\end{center}

Массовую скорость, отвечающую непрерывному спектру, разложим по
степеням коэффициента диффузности:
$$
U_c(x)=U_c^{(0)}(x)+qU_c^{(1)}(x)+q^2U_c^{(2)}(x)+\cdots.
\eqno{(5.2)}
$$

Тогда профиль массовой скорости в полупространстве можно строить
по формуле:
$$
U(x)=U_{sl}(q)+G_vx+U_c(x),
$$
где $U_c(x)$ определяется равенством (5.2).

Коэффициенты ряда (5.2) вычислим согласно выведенным выше
формулам:
$$
U_c^{(n)}(x)=G_v\dfrac{2-q}{2\pi}\int\limits_{-\infty}^{\infty}
e^{ikx}E_n(k)dk, \qquad n=0,1,2,\cdots .
$$

Вычислим скорость газа непосредственно у стенки:
$$
U(0)=U_{sl}(q)+U_c^{(0)}(0)+qU_c^{(1)}(0)+q^2U_c^{(2)}(0)+\cdots.
\eqno{(5.3)}
$$

В случае чисто диффузного отражения молекул от стенки ($q=1$)
согласно (5.3) мы имеем
$$
U(0)=U_{sl}(1)+U_c^{(0)}(0)+U_c^{(1)}(0)+U_c^{(2)}(0)+\cdots.
$$

Отсюда в нулевом приближении получаем:
$$
U^{(0)}=U_{sl}(1)+U_c^{(0)}(0)=0.4382G_v.
$$
В первом приближении получаем:
$$
U^{(1)}(0)=U_{sl}(1)+U_c^{(0)}(0)+U_c^{(1)}(0)=0.4482G_v,
$$

Сравним эти результаты с точным значение скорости у стенки
(\cite{93}, \cite{LY-2})
$$
U(0)=\dfrac{1}{\sqrt{5}}G_v=0.4472G_v.
$$

Относительная ошибка
$$
O_n=\dfrac{U(0)-U^{(n)}(0)}{U(0)}\cdot 100\% , \qquad
n=0,1,2,\cdots,
$$
в нулевом приближении равна $-2.01\%$, а уже в первом
приближении составляет менее $-0.22\%$.

Аналогично скорости скольжения, получим профиль размерной
массовой скорости в полупространстве. Исходя из того, что
$$
U_y(x)=V_1G_v+G_vx +
G_v\dfrac{2-q}{2\pi}\int\limits_{-\infty}^{\infty}
e^{ikx}E_n(k)dk,
$$
или, иначе:
$
{U_y(x)}=H(x,\alpha)G_v,
$
где
$$
H(x,\alpha)=V_1+x+\dfrac{2-q}{2\pi}\int\limits_{-\infty}^{\infty}
e^{ikx}E_n(k)dk,
$$
Теперь получаем следующее выражение для размерной массовой скорости
газа:
$$
u_y(x)=\dfrac{U_y(x)}{\sqrt{\beta}}=
\dfrac{H(x,\alpha)}{\sqrt{\beta}}G_v(\alpha)=
\dfrac{H(x,\alpha)}{\sqrt{\beta}l}G_v(\alpha)l.
$$
Выполняя преобразования, аналогичные тем, которые использовались
для получения размерной скорости скольжения, приходим к следующему выражению
для размерной массовой скорости:
$$
u_y(x)=K_{v}^{*}(x, \alpha)l g_v,
$$
где
$$
K_{v}^{*}(x,\alpha)=\dfrac{15H(x,\alpha)l_0(\alpha)}{8\sqrt{\pi}
l_1(\alpha)}.
$$

На рисунках 1, 2 и 3 построены профили массовой скорости для
различных значений коэффициента диффузности. Отметим, что
химический потенциал молекул $\alpha$ принимается равным $-5$.

\begin{figure}[h]
\begin{center}
\includegraphics[width=14.0cm, height=6.5cm]{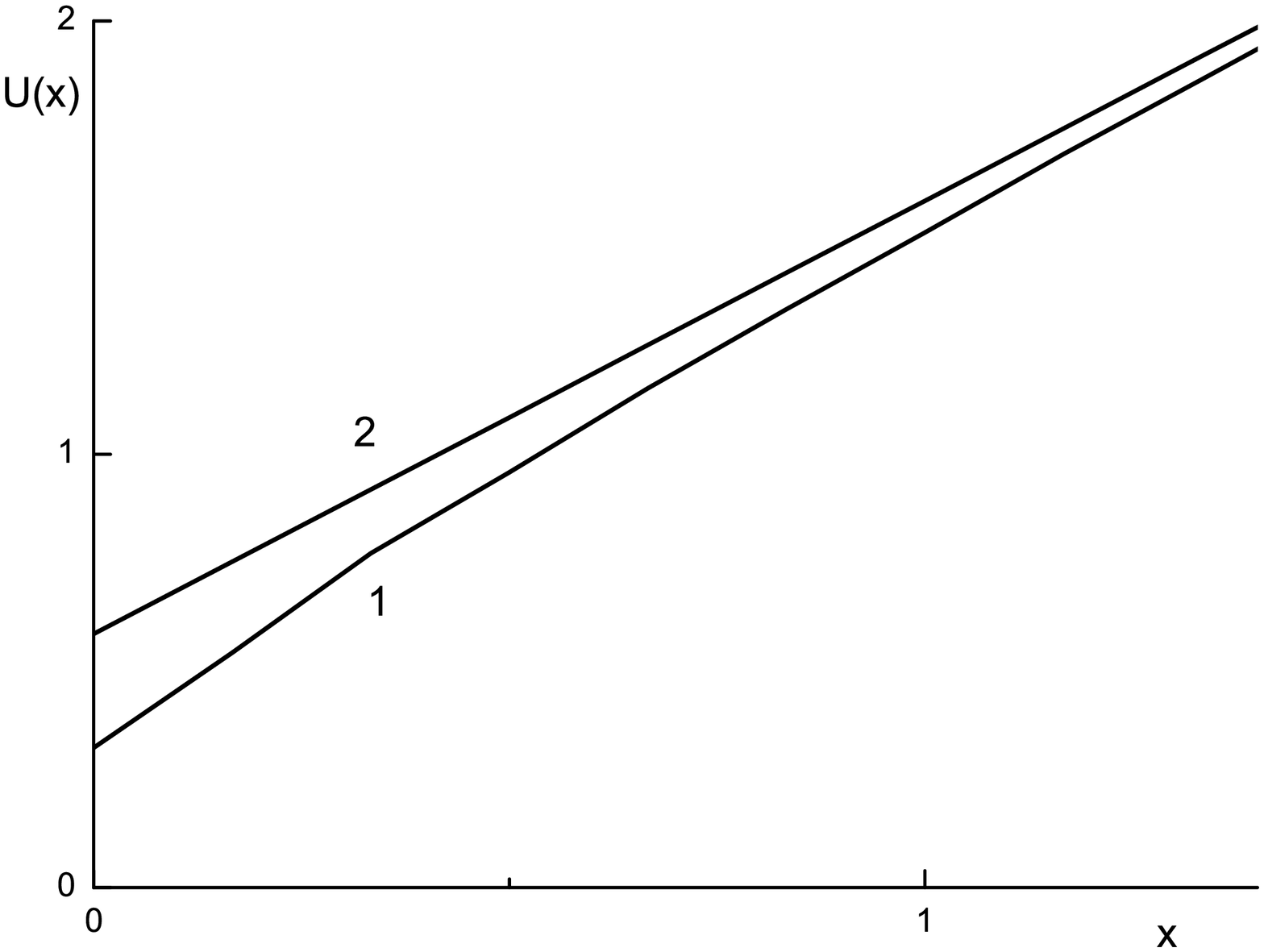}

Рис. 1. Профиль массовой скорости в полупространстве,
коэффициент диффузности равен: $q=1$, химический потенциал
молекул $\alpha=-5$.
\end{center}
\end{figure}

\begin{figure}[hb]
\begin{center}
\includegraphics[width=14.0cm, height=6.5cm]{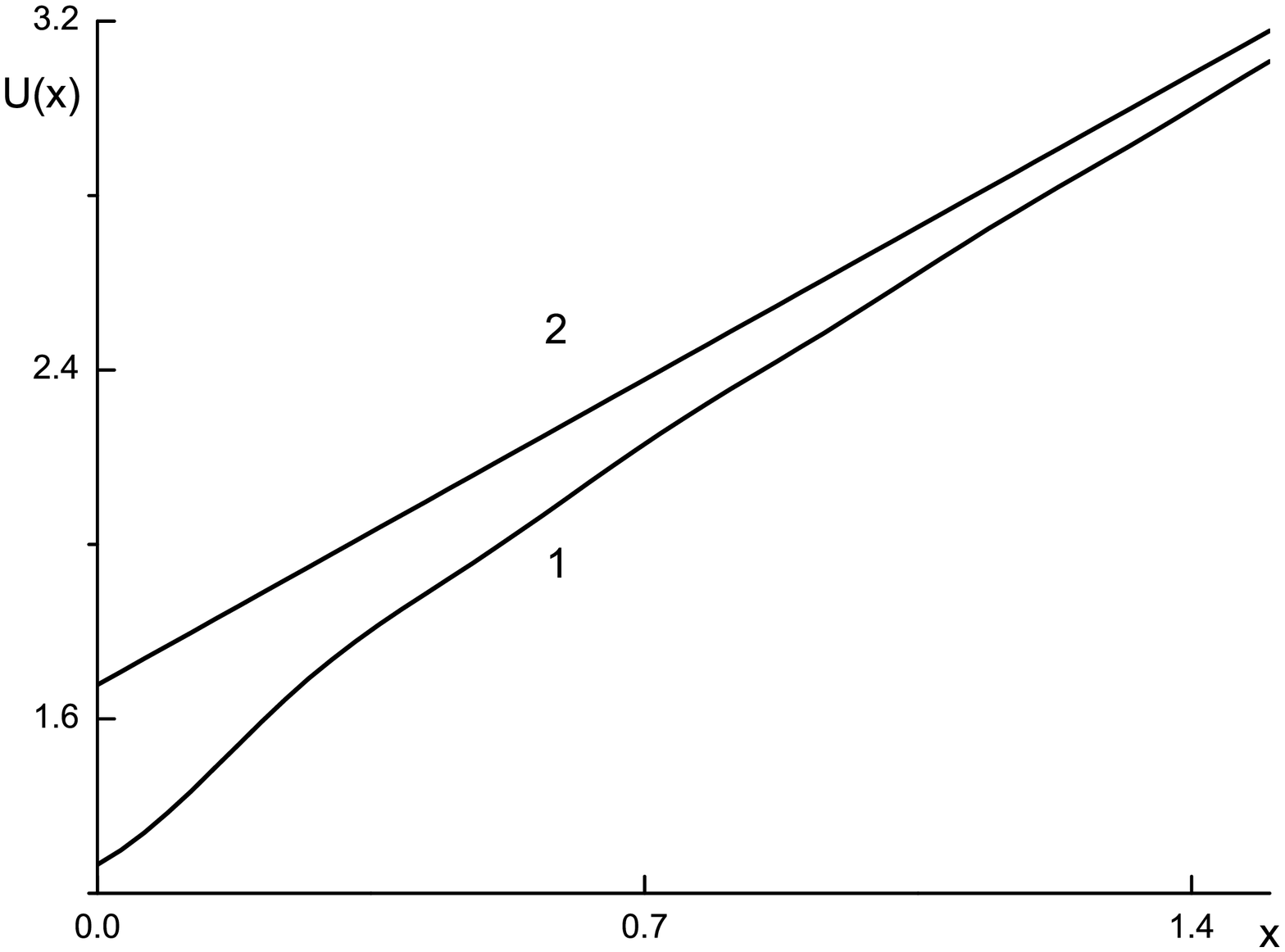}

Рис. 2. Профиль массовой скорости в полупространстве,
коэффициент диффузности равен: $q=0.5$, химический потенциал
молекул $\alpha=-5$.
\end{center}
\end{figure}

\newpage

\begin{figure}
\begin{center}
\includegraphics[width=14.0cm, height=7.5cm]{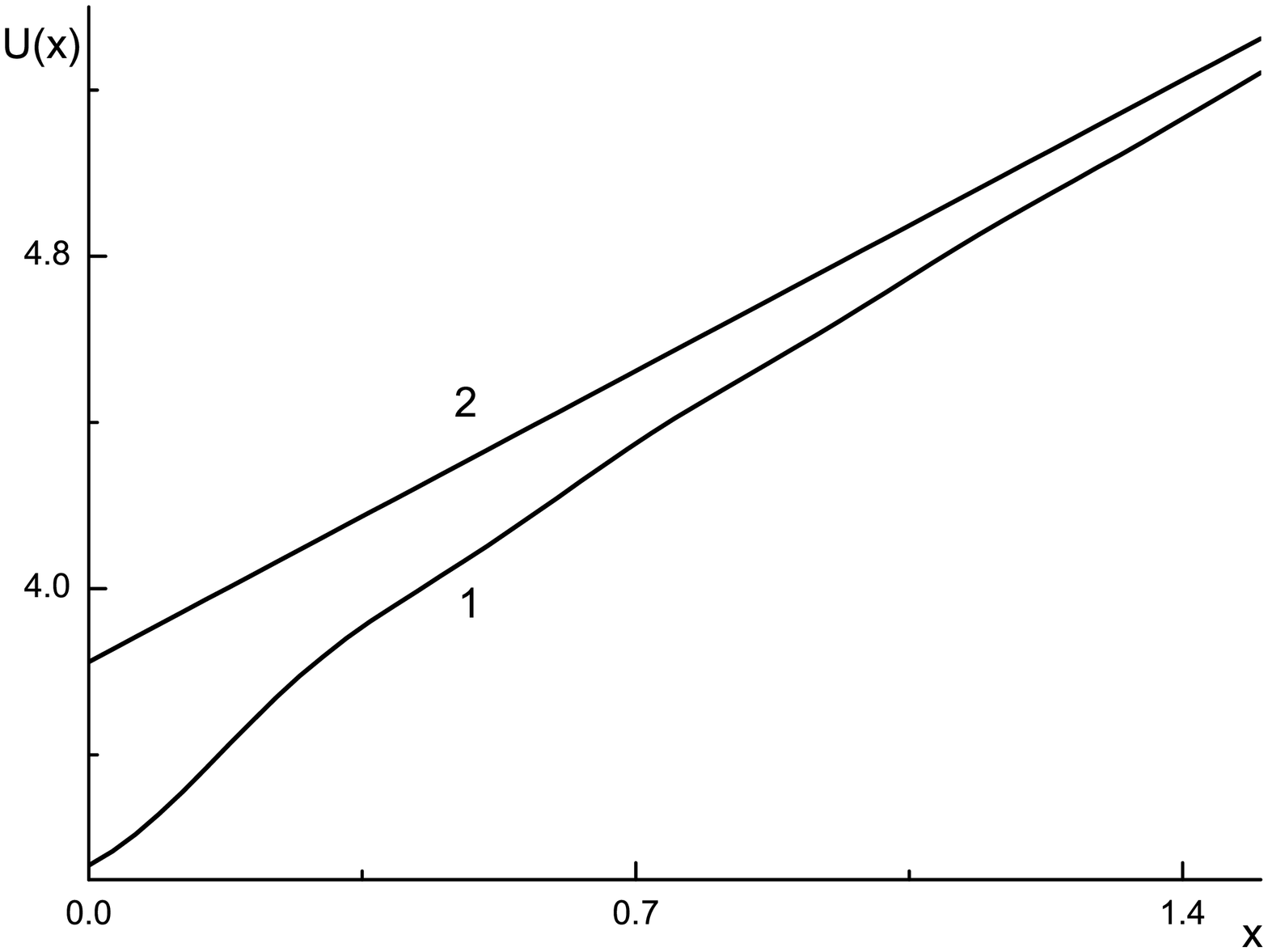}

Рис. 3. Профиль массовой скорости в полупространстве,
коэффициент диффузности равен: $q=0.25$, химический потенциал
молекул $\alpha=-5$.
\end{center}
\end{figure}

\renewcommand{\baselinestretch}{1.03}

\end{document}